\documentclass{elsart}

\usepackage{epsfig,rotating}

\usepackage{amssymb}

\newcounter{figcount}

\begin{document}

\pagestyle{plain}

\begin{frontmatter}



\title{Performance of ALICE silicon pixel detector \\
prototypes in high energy beams}

\thanks[talk]{
Corresponding author, {\em E-mail address: Domenico.Elia@ba.infn.it}
\\
on behalf of the SPD project in the 
ALICE Collaboration}


%\author{
D.~Elia$^{a,\star}$,
G.~Anelli$^{b}$,
F.~Antinori$^{c}$,
A.~Badal\`a$^{d}$,
G.E.~Bruno$^{a}$,
M.~Burns$^{b}$,
I.A.~Cali$^{a,b}$,
M.~Campbell$^{b}$,
M.~Caselle$^{a}$,
S.~Ceresa$^{b}$,
P.~Chochula$^{b}$,
M.~Cinausero$^{e}$,
J.~Conrad$^{b}$,
R.~Dima$^{c}$,
D.~Fabris$^{c}$,
R.A.~Fini$^{a}$,
E.~Fioretto$^{e}$,
S.~Kapusta$^{b}$,
A.~Kluge$^{b}$,
M.~Krivda$^{f}$,
V.~Lenti$^{a}$,
F.~Librizzi$^{d}$,
M.~Lunardon$^{c}$,
V.~Manzari$^{a}$,
M.~Morel$^{b}$,
S.~Moretto$^{c}$,
P.~Nilsson$^{b}$,
F.~Osmic$^{b}$,
G.S.~Pappalardo$^{d}$,
V.~Paticchio$^{a}$,
A.~Pepato$^{c}$,
G.~Prete$^{e}$,
A.~Pulvirenti$^{d}$,
P.~Riedler$^{b}$,
F.~Riggi$^{d}$,
L.~S\'andor$^{f}$,
R.~Santoro$^{a}$,
F.~Scarlassara$^{c}$,
G.~Segato$^{c}$,
F.~Soramel$^{g}$,
G.~Stefanini$^{b}$,
C.~Torcato de Matos$^{b}$,
R.~Turrisi$^{c}$,
L.~Vannucci$^{e}$,
G.~Viesti$^{c}$,
T.~Virgili$^{h}$
%}
%\vspace{-3mm}

\address[a]{Dipartimento IA di Fisica dell'Universit{\`a} and INFN, 
Bari, Italy}
\vspace{-1mm}
\address[b]{CERN, Geneva, Switzerland}
\vspace{-1mm}
\address[c]{Dipartimento di Fisica dell'Universit{\`a} and INFN,
Padova, Italy}
\vspace{-1mm}
\address[d]{Dipartimento di Fisica dell'Universit{\`a} and INFN,
Catania, Italy}
\vspace{-1mm}
\address[e]{Laboratori Nazionali di Legnaro, Legnaro, Italy}
\vspace{-1mm}
\address[f]{Institute of Experimental Physics, Slovak Academy of Science,
Ko\v{s}ice, Slovakia}
\vspace{-1mm}
\address[g]{Dipartimento di Fisica dell'Universit{\`a} and INFN, 
Udine, Italy}
\vspace{-1mm}
\address[h]{Dipartimento di Fisica dell'Universit{\`a} and INFN,
Salerno, Italy}

\begin{abstract}
The two innermost layers of the ALICE inner tracking system
are instrumented with silicon pixel detectors.
Single chip assembly prototypes of the ALICE pixels
have been tested in high energy particle beams at the CERN SPS.
Detection efficiency and spatial precision have been studied 
as a function of the threshold 
and the track incidence angle.
The experimental method, data analysis and main
results are presented. 
\end{abstract}

\begin{keyword}
Spatial precision \sep Efficieny \sep
Silicon pixel detector \sep ALICE \sep LHC
\PACS 29.40.Gx \sep 29.40.Wk
\end{keyword}
\end{frontmatter}

% main text

\section{Introduction}
The ALICE experiment is dedicated to the study of
the properties of QCD matter created in
nucleus-nucleus collisions at LHC energies~\cite{TechProp}.
The inner tracking system in the ALICE apparatus 
is made of position sensitive detectors which
have to handle 
several thousands tracks per unit of rapidity.
The two innermost layers 
at 3.9 $cm$ and 7.6 $cm$ radii, respectively, constitute 
the Silicon Pixel Detector (SPD).
The spatial precision and hit efficiency of the SPD 
are key parameters since they determine 
the ALICE capability of detecting particles with open 
heavy-flavour~\cite{ITStdr}. 
\\
The basic detector unit of the ALICE SPD is the ladder, 
a two-dimensional silicon matrix of p$^+$n 
reverse biased diodes of dimensions 50~x~425~$\mu{m}^2$, 
flip-chip bonded to five 
read-out chips. Each diode is connected to a cell of the front-end 
read-out ASIC via a Pb-Sn solder bump of 25~$\mu{m}$ diameter. 
The detector contains nearly 10$^7$ active cells in total. 
The read-out is binary. 
To reduce the material budget, the sensor thickness is 
limited to 200 $\mu{m}$ and the read-out chip wafers
are thinned down to 150 $\mu{m}$.
Further details can be found 
in \cite{PetraPix2005}.
\\
Early prototypes of the ALICE SPD elements, in the form
of single-chip assemblies, were tested in high energy 
proton/pion beams at the CERN SPS in 2002 and 2003.
These assemblies were made with sensors of
200 $\mu{m}$ and 300 $\mu{m}$ thicknesses, while the
read-out chips (unthinned) were 725 $\mu{m}$ thick.
Those beam tests were primarily aimed at evaluating the 
performance of the bump-bonded assemblies and of the read-out 
electronics. Spatial precision and hit efficiency studies
would have in principle required a more precise 
tracking telescope. However we found that the performance of the 
tracking doublets, together with a detailed cluster analysis of the hits, 
can yield a good determination of the intrinsic spatial 
precision and detection efficiency of the pixel plane under test.
In the following sections the main focus is on 
the results of the 2002 beam test, where the sensor thickness (200 $\mu{m}$) was the same as 
the one used in ladder production; a study of
the detector performance as 
a function of the threshold and the track incidence
angle is presented.
Some comparisons with the main results
for the thicker sensor are also discussed.

\section{Experimental setup and data analysis} 

\begin{figure}[htb]
\centering
\includegraphics[scale=0.5]{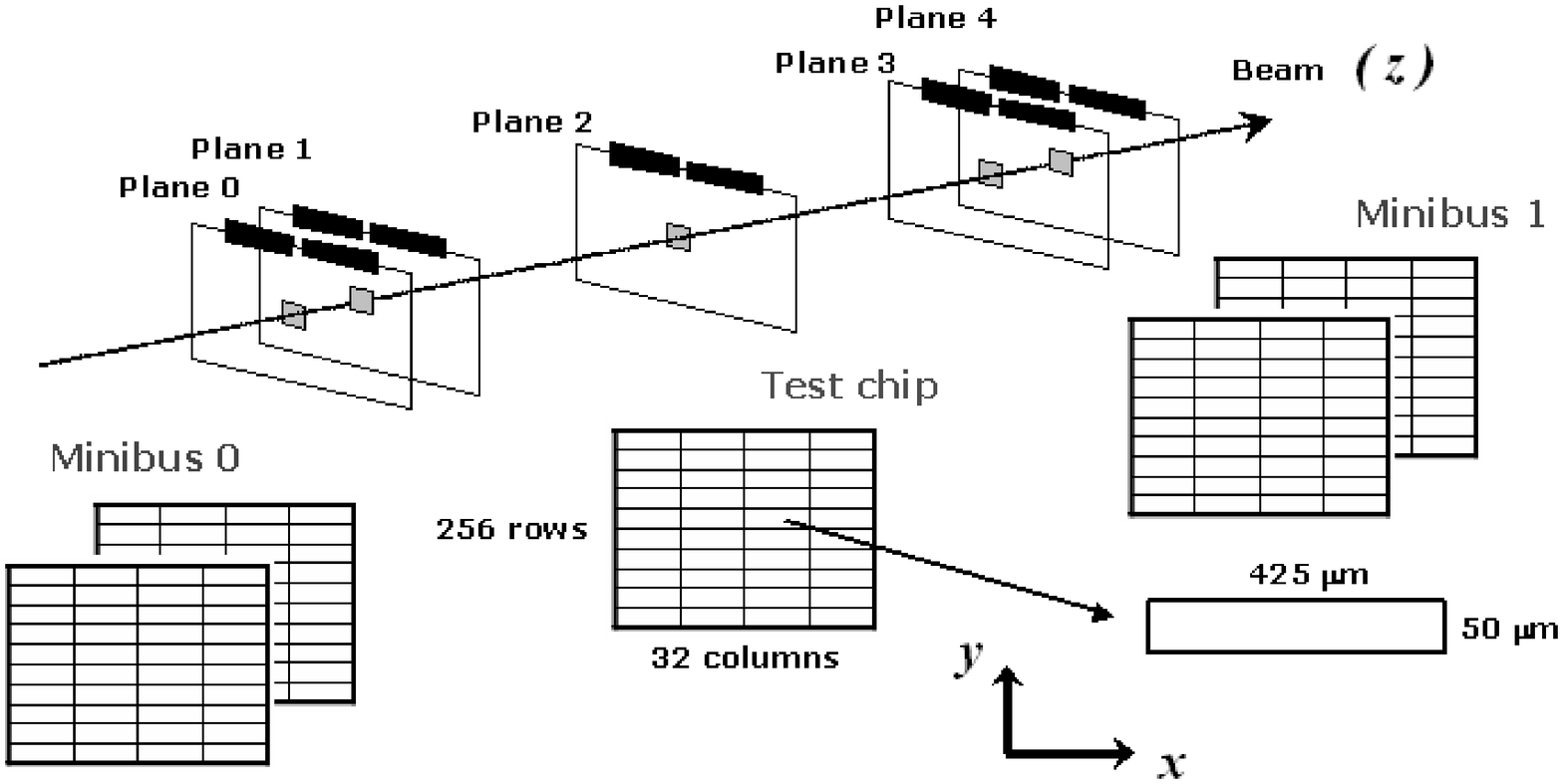}
\caption{}
\label{figsetup2002fine}
\end{figure}

\begin{figure}[hb!]
\centering
\includegraphics[scale=0.5]{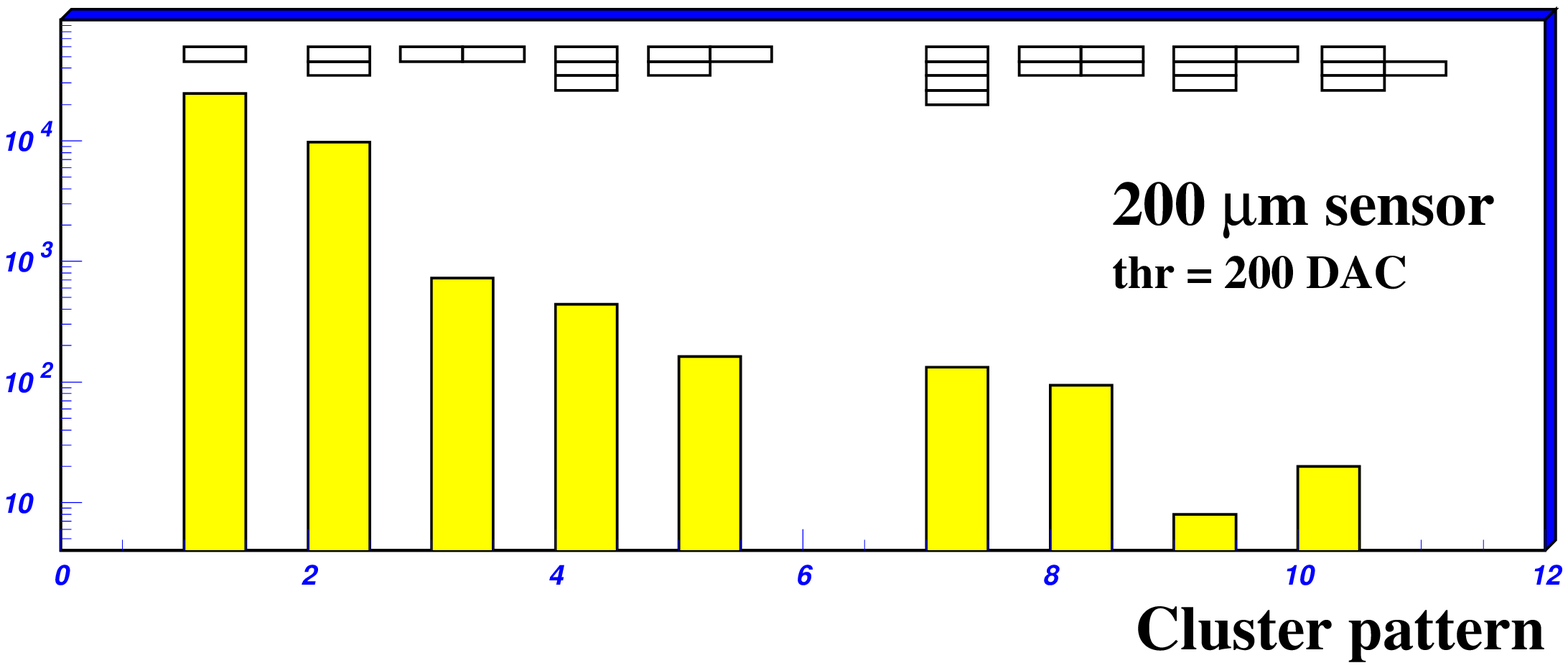}
\caption{}
\label{figclustertype02_tilt0_th200_log}
\end{figure}

The 2002 test was carried out 
in the H4 beam line at the SPS with a 350 GeV/$c$ proton 
beam~\cite{PetraPixel02}.
The experimental setup is schematically 
shown in Fig.\ref{figsetup2002fine}.
The assembly under test was placed between two assembly pairs
(minibus doublets) that were used as tracking telescope.
The assemblies were all positioned with the short (50 $\mu{m}$) 
pixel cell side aligned along the $y$ axis.
The transverse position and the tilt angle of the test plane with 
respect to the beam line 
could be changed remotely using a stepping motor gearbox.
The trigger signal was generated by 
a telescope of four scintillating counters.
\\
Data was collected with different inclination angles
of the test plane with respect to the beam line,
corresponding to rotations along an axis parallel
to the 425 $\mu{m}$ cell size direction
($x$ coordinate).
An angle scan (from 0 to 30, in steps of 5 degrees)
was performed for three different values of the threshold,
around the typical operating setting
(namely 185, 200 and 210 DAC units).\footnote{The DAC value of 
200 corresponds to about 3,000 electrons
in the effective threshold.
Each 15 units decrease in the DAC setting corresponds
to about 1,000 electrons increase in the effective
threshold, at least for DAC values above 150.}
A complete scan of the global detector threshold was performed 
only for the configuration with normal incidence 
angle tracks.
\\
In 2003 another test was done  
in the same experimental area, with a 120 GeV/$c$ proton/pion beam as well as
with higher multiplicity secondaries from
158 $A$ GeV/$c$ indium ions hitting a lead target in 
front of the detector setup.
In this case the detector under study was an assembly
with a 300 $\mu{m}$ thick sensor
and the two tracking detectors in each minibus doublet
had the short and the long pixel cell side, respectively, 
aligned to the $y$ axis: this crossed geometry was 
adopted to optimize the measurement of the transverse position 
of the tracks in both $x$ and $y$ coordinates.
For more details see \cite{PaulNilssonVienna04,AlbertoPix2005}. 
\\
The analysis method is mainly based on the following steps:
the cluster analysis of the hits on the pixel planes, the alignment of the
transverse position of the detectors, the beam track
reconstruction by using the tracking doublets and the study
of the residual distributions on the test plane:
a detailed description can be found in \cite{Notatb2003,Notatb2002}.
Clusters on the test plane are assigned to the beam track
when their distance from the track impact prediction
is below a defined maximum (i.e a cut in the residual
is applied).
In Fig.\ref{figclustertype02_tilt0_th200_log} the frequency of different
cluster patterns, at 200 DAC threshold
($\approx$~3,000 electrons) and for
normal incidence tracks,
is shown.
Under these conditions most of the clusters are made by a single pixel (68\%) or
two pixels along the direction of the short cell side (27\%), 
while patterns with more  
than two pixels along the 425 $\mu{m}$ cell side are not found.
The relative frequencies of the various cluster types, as shown in the following
section, rapidly change with the threshold 
and the track incidence angle on the detector since they are
mostly determined by charge sharing and
geometrical effects.
\\
In absence of charge sharing, all the charge carriers 
locally generated around the incident particle trajectory
are collected on a single pixel cell. The spatial 
precision in this case is given by
$L/\sqrt{12}$ (where
$L$ is the pitch size of the detection 
element), which for the ALICE SPD corresponds to:

\begin{equation}
\sigma_{pixel}^{ncs}(y) = 14.4 \hspace{2mm} \mu{m} 
\hspace{15mm}
\sigma_{pixel}^{ncs}(x) = 122.7 \hspace{2mm} \mu{m} 
\hspace{35mm}.
\label{eqthprecisionvalues}
\end{equation}
\vspace{-5mm}

In order to obtain the intrinsic spatial precision
of the detector, the tracking error has to be subtracted from the
widths of the residual distributions.
As an approximation
we have assumed the following expression:

\begin{equation}
\sigma_{pixel}^2(y) = \sigma_{resid}^2(y) - \sigma_{track}(y)^2  
\label{eqmeasprecision}
\end{equation}
\vspace{-5mm}

\noindent
where $\sigma_{resid}(y)$ is the sigma of the Gaussian fit 
to the measured $y$-residual distributions, while
$\sigma_{track}(y)$ has been estimated to be about 
6 $\mu{m}$~\cite{Notatb2002}.
\\
The efficiency of the test plane is
defined as the ratio between the number of events 
where a cluster correlated to the reconstructed track 
is detected, 
and the total number of reconstructed tracks. 
The efficiency calculation has been based on a large number of 
tracks and cross checked for uniformity
across the whole detector surface.

\section{Results}

\begin{figure}[hb!]
\centering
\includegraphics[scale=0.5]{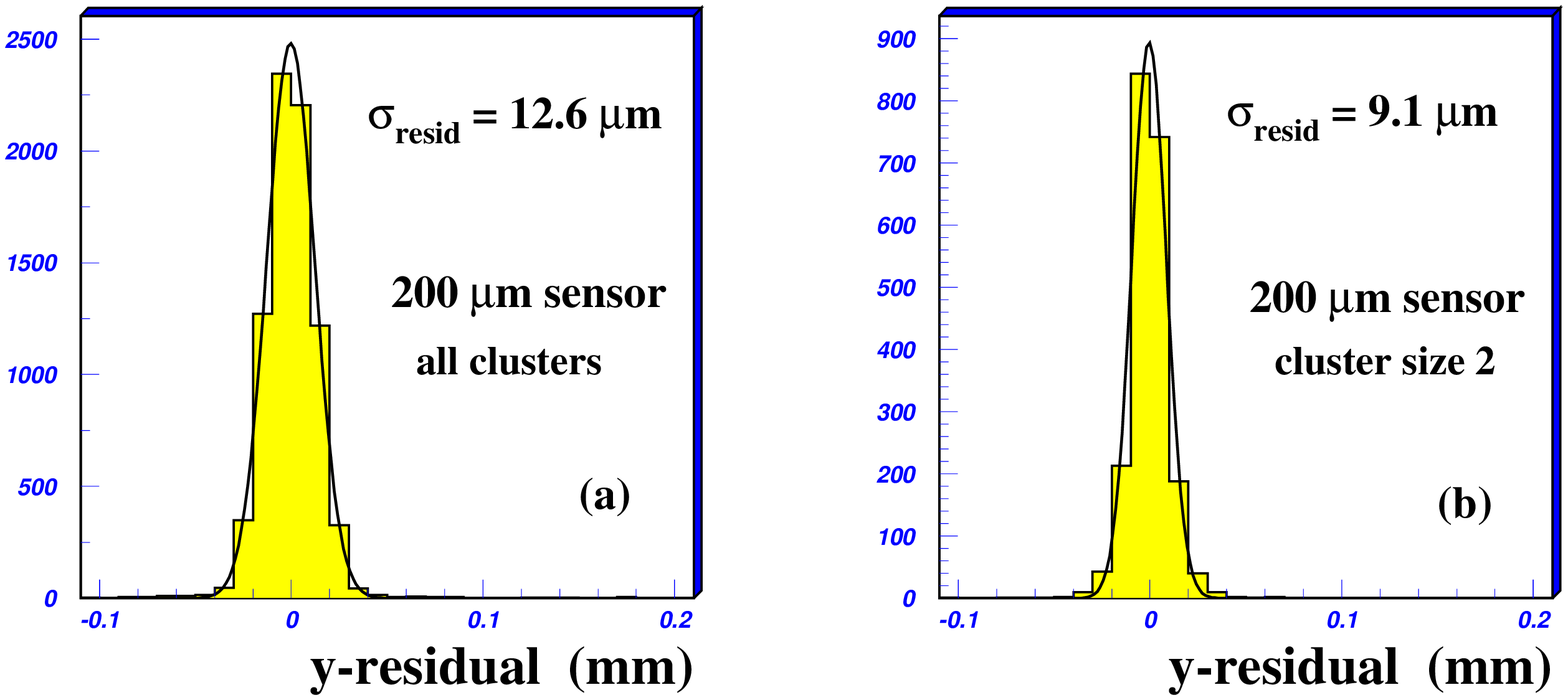}
\caption{}
\label{figresiduals02allcls2}
\end{figure}

% As already shown, 
At normal track incidence and 200 DAC threshold
mainly single pixel and double-$y$ pixel clusters occur.
The precision is determined by their relative abundance.
In Fig.\ref{figresiduals02allcls2}(a) the total $y$-residual
distributions (for all the cluster patterns) is shown.
After subtracting the uncertainty on the telescope prediction,
the global detector precision
is found to be
$\sigma_{pixel}(y) =
(\hspace{1mm} 11.1 \hspace{1mm} \pm \hspace{1mm} 0.2 
\hspace{1mm}) \hspace{2mm} \mu{m}$.
\\
Due to charge sharing effects, a particle crossing a pixel cell
% within a small region from the border 
% (charge sharing region)
close to the edge may generate two-pixel clusters 
with improved precision in the impact localization. 
This explains the narrower double-$y$ pixel cluster residual 
distribution shown in Fig.\ref{figresiduals02allcls2}(b).
The spatial precisions corresponding to the two main cluster patterns 
separately are found to be:

\begin{equation}
\sigma_{pixel}^{cls 1}(y) =
(\hspace{1mm} 11.5 \hspace{1mm} \pm \hspace{1mm} 0.2 
\hspace{1mm}) \hspace{2mm} \mu{m} \hspace{11mm}
\sigma_{pixel}^{cls 2}(y) =
(\hspace{1mm} 6.8 \hspace{1mm} \pm \hspace{1mm} 0.3 
\hspace{1mm}) \hspace{2mm} \mu{m}
\hspace{8mm}.
\label{eqmeasprecisionvalues2002}
\end{equation}
\vspace{-4mm}

\noindent
The errors on these estimates
have been calculated by taking into account both the
contributions due to uncertainties in the widths of 
the residual distributions and uncertainty in the 
tracking precision. 
A similar study for the spatial precision along the long cell
side coordinate is available only for the
300 $\mu{m}$ sensor~\cite{Notatb2003}.
At a DAC setting of 200 the efficiency has been estimated to be above 99\%.
In the following sections the performance dependence on 
the detector threshold and the track incidence
angle are discussed.

\subsection{Detector performance as a function of the threshold}
\begin{figure}[hb!]
\centering
\includegraphics[scale=0.55]{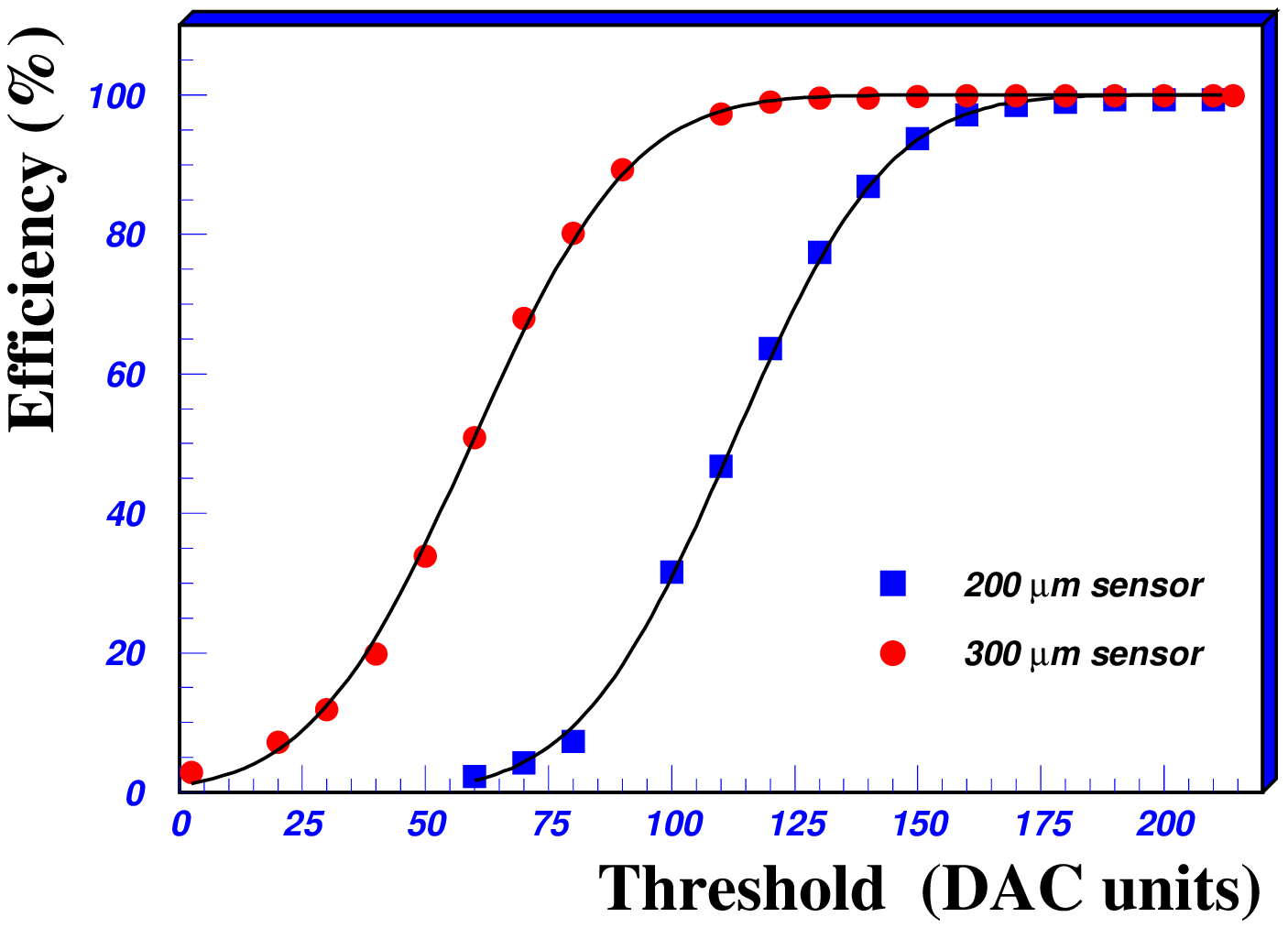}
\caption{}
\label{figefficiencyvsthr02}
\end{figure}

\begin{figure}[hb!]
\centering
\includegraphics[scale=0.6]{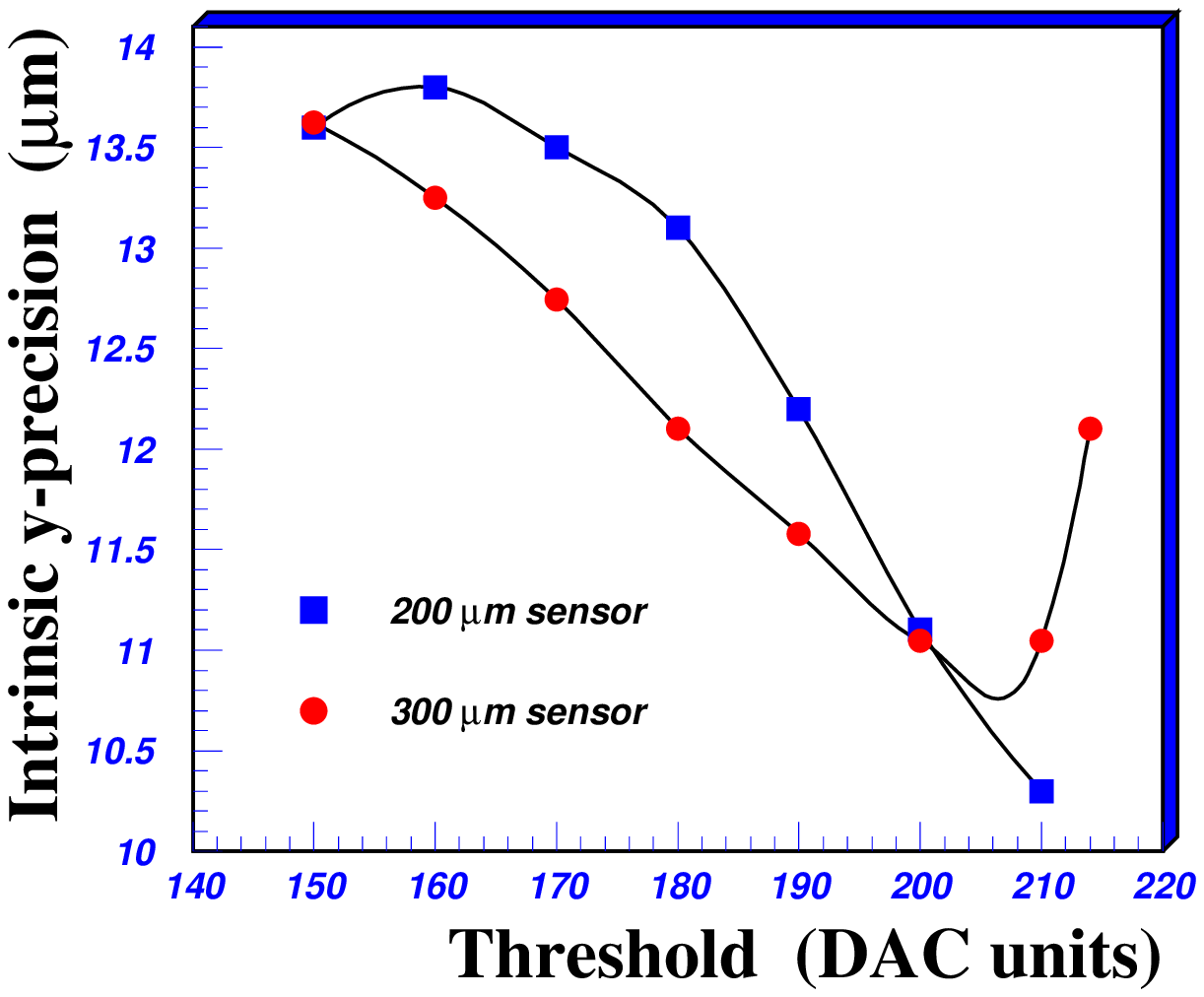}
\caption{}
\label{figresolutionsvsthr02}
\end{figure}

\begin{figure}[hb!]
\includegraphics[scale=0.5]{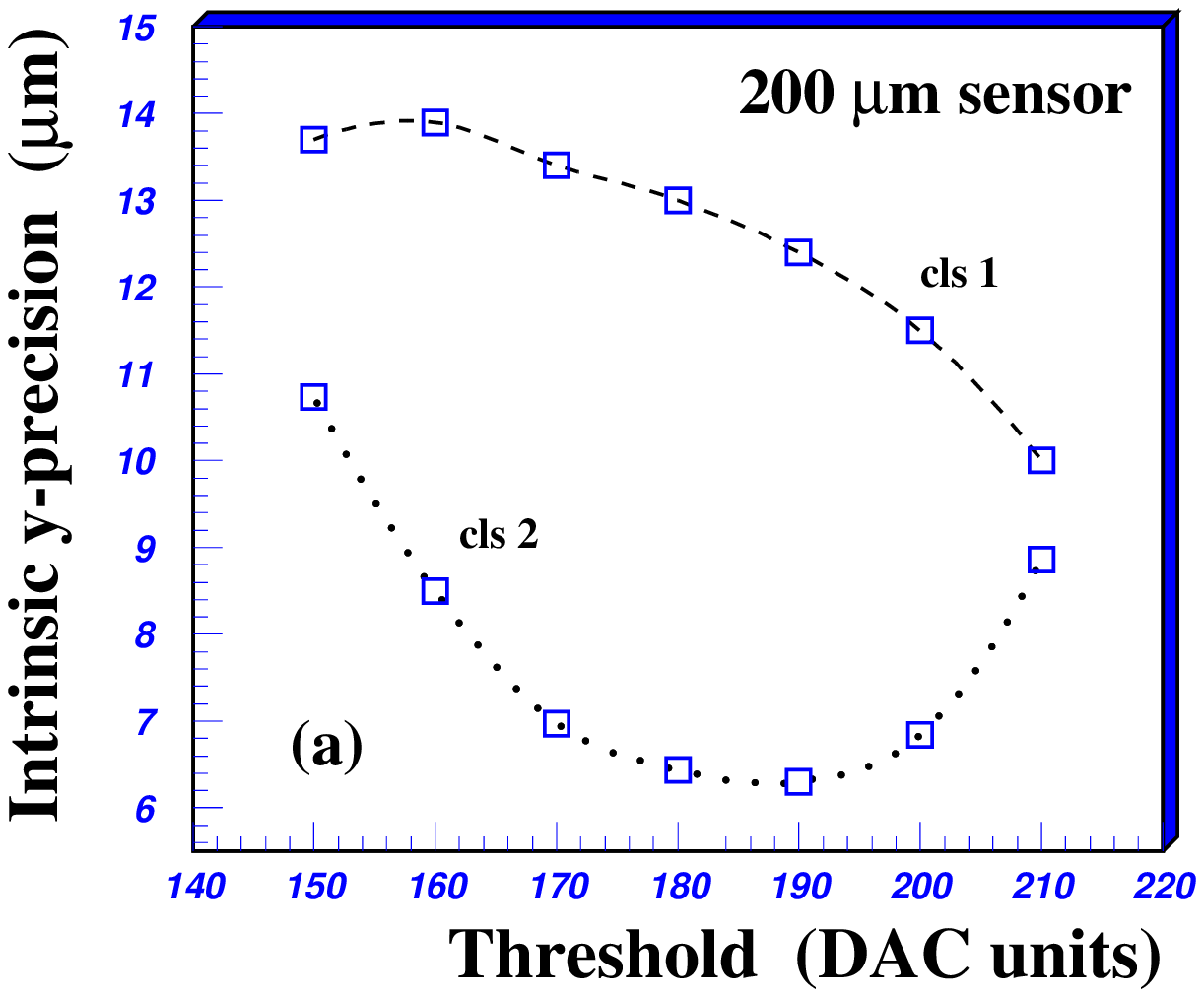}
\hspace{\fill}
\begin{minipage}[t]{110mm}
\includegraphics[scale=0.5]{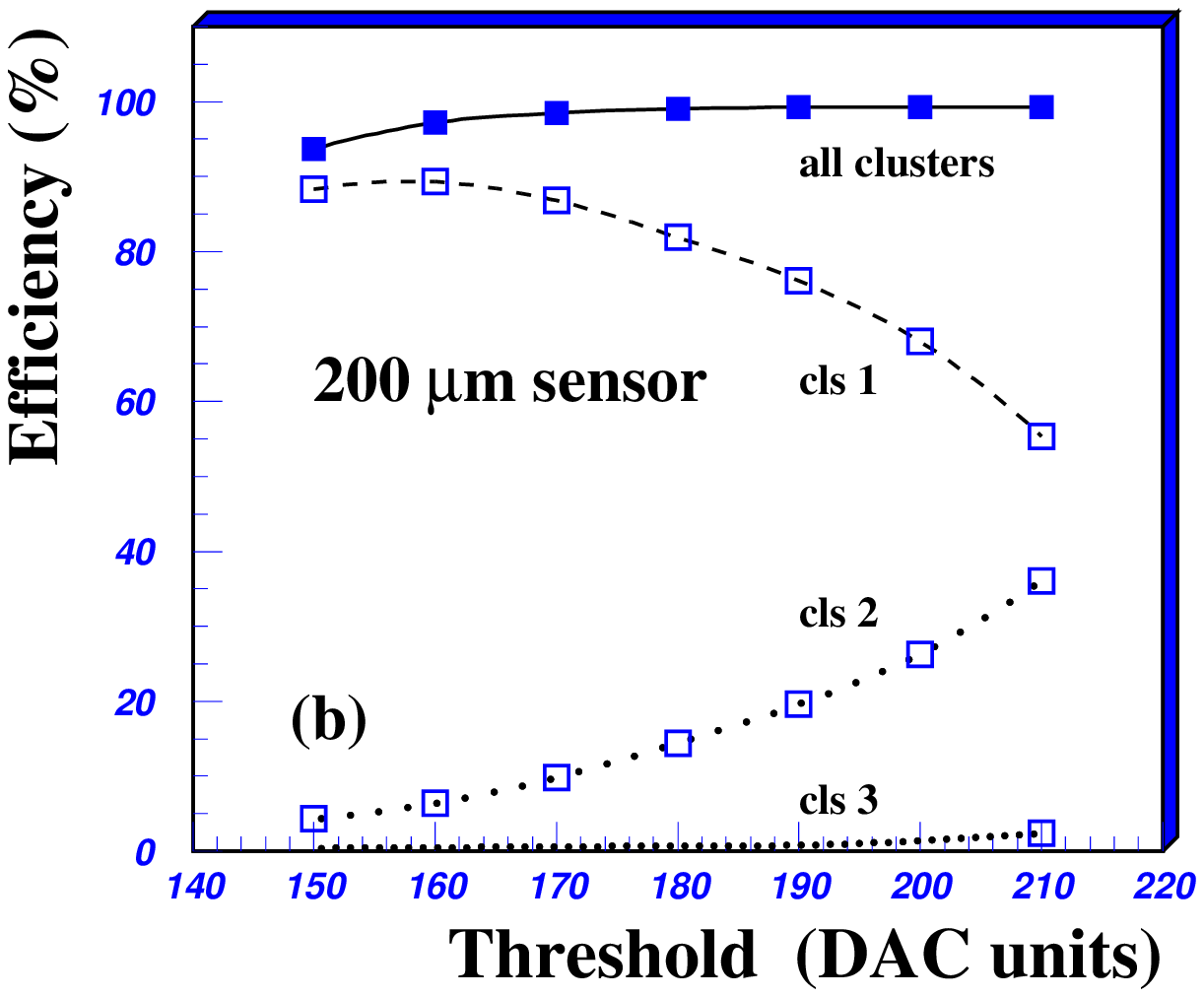}
\end{minipage}
\caption{}
\label{figresolutionsefficvsthr02_cl1cl2}
\end{figure}

A complete threshold scan measurement was performed with tracks
at normal incidence.
A detailed study of the evolution of the detector response,
in particular for the relative frequencies of the
various cluster patterns,
has been carried out and used to adjust 
the SPD simulation~\cite{Notatb2002,NotaPeppe}.
\\
The dependence of the detector efficiency
on the threshold is illustrated in Fig.\ref{figefficiencyvsthr02},
where results are compared with those for the
300 $\mu{m}$ sensor assembly. 
The detectors are fully
efficient in a wide plateau region (efficiency values above 99\%). 
Due to the smaller amount of produced charge carriers,
the values for the thin sensor detector are
shifted towards lower thresholds, hence
the plateau region itself is shorter.
% The superimposed curves
% are the result of the fit to the
% Gaussian-integral
% error function.
\\
The dependence of the spatial precision 
on the global threshold is shown in 
Fig.\ref{figresolutionsvsthr02}, again compared with
the corresponding result found for the 300 $\mu{m}$ sensor.
A steeper dependence on the threshold for the thin detector is found.
The smooth lines are from a spline fitting algorithm.
The absolute minimum in the precision for the 300 $\mu{m}$ sensor
is reached
% , as expected,
at the threshold for which single
and double-$y$ pixel clusters are equally frequent~\cite{Notatb2003}.
This condition for the thin sensor is not
even reached at lowest threshold of 210 DAC ($\approx$ 2,000 electrons):
this is shown in Fig.\ref{figresolutionsefficvsthr02_cl1cl2}
where the contributions of single and double-$y$ pixel
clusters to the intrinsic precision (a) and to the detection 
efficiency (b) are illustrated.
\\
The thin detector precision curve reaches a maximum
around 160 DAC setting, with a corresponding value
closely approaching $L_y/\sqrt{12}$ (mostly single pixel clusters).
Tracks impacting a cell close enough to the boundary region
may share the produced charge almost equally in two
adjacent pixels: for 
%very
high thresholds (DAC values smaller than 160
for the 200 $\mu{m}$ sensor case) it can happen that
none of the two pixels are fired. For this reason below
160 DAC the sensitive region of the pixel cell is reduced:
this explains the
decrease of the intrinsic precision which appears
at the same threshold setting where the
efficiency also starts to decrease with respect to the plateau value.
\\
The width of the charge sharing region 
increases for decreasing thresholds: this explains
the increase of the precision curve for double
pixel clusters when increasing the DAC setting.
Below 190 DAC units ($\approx$ 4,000 electrons) mostly cluster patterns
% topologies
that would have been made by 3 or 4 pixels for softer thresholds enter the double-$y$ pixel cluster sample, then 
making the corresponding average precision slightly worse.
In addition, at high thresholds, 
the fraction of double pixel clusters created by charge sharing 
are expected to be negligible with respect those originated, for instance,
from delta-rays.

\subsection{Detector performance as a function of the track incidence angle}
\begin{figure}[hb!]
\hspace{20mm} 
\includegraphics[scale=0.35]{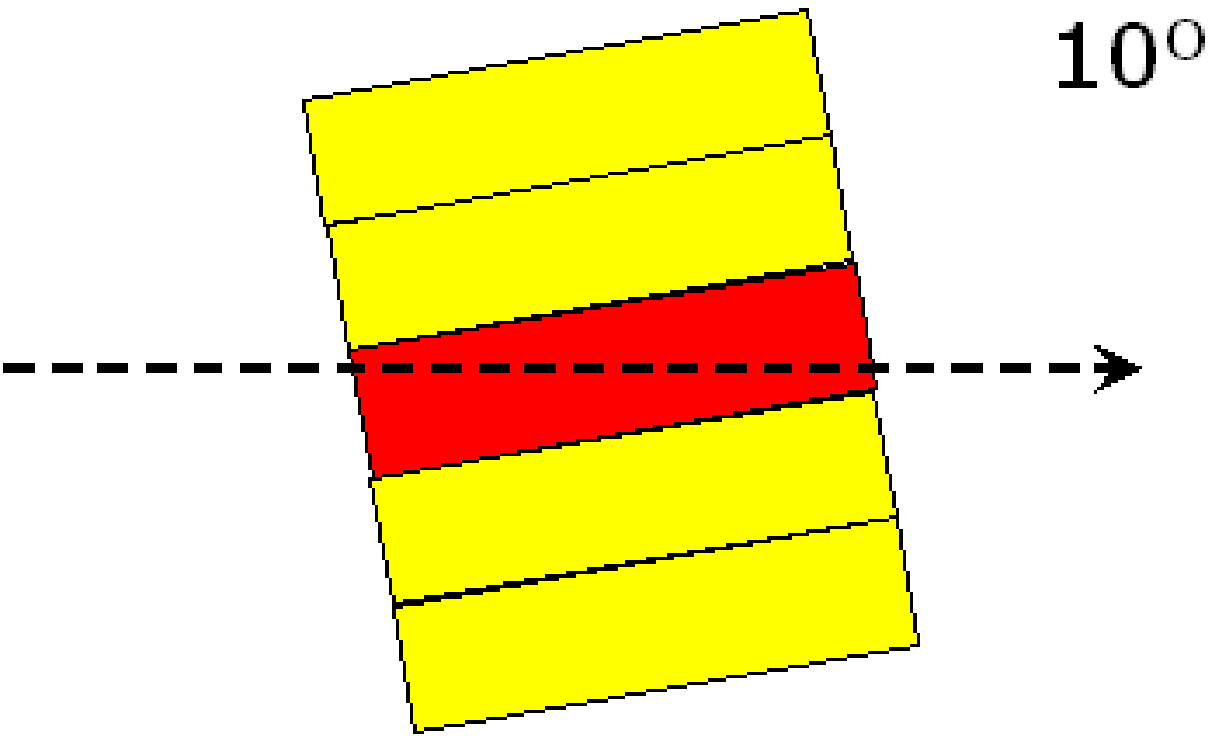}
\hspace{10mm} 
\hspace{\fill}
\begin{minipage}[t]{110mm} 
\includegraphics[scale=0.35]{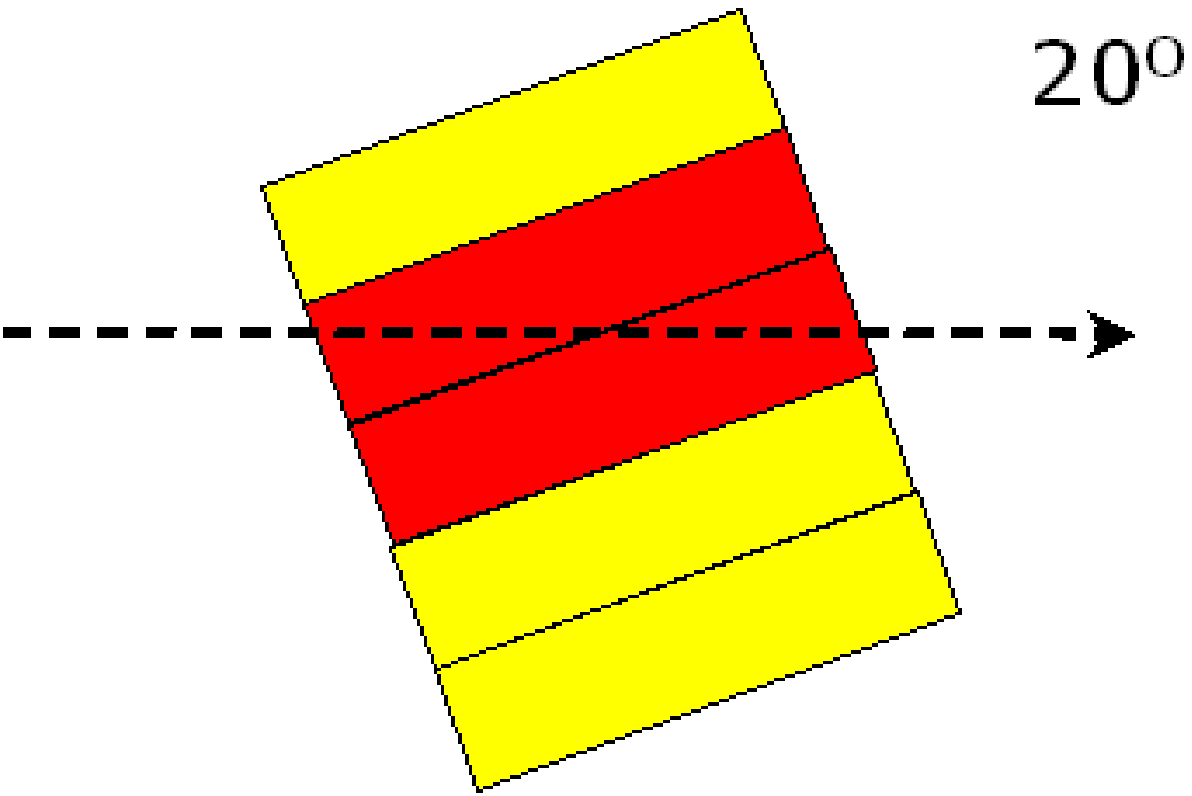}
\end{minipage} 
\\
\begin{center}   
\includegraphics[scale=0.35]{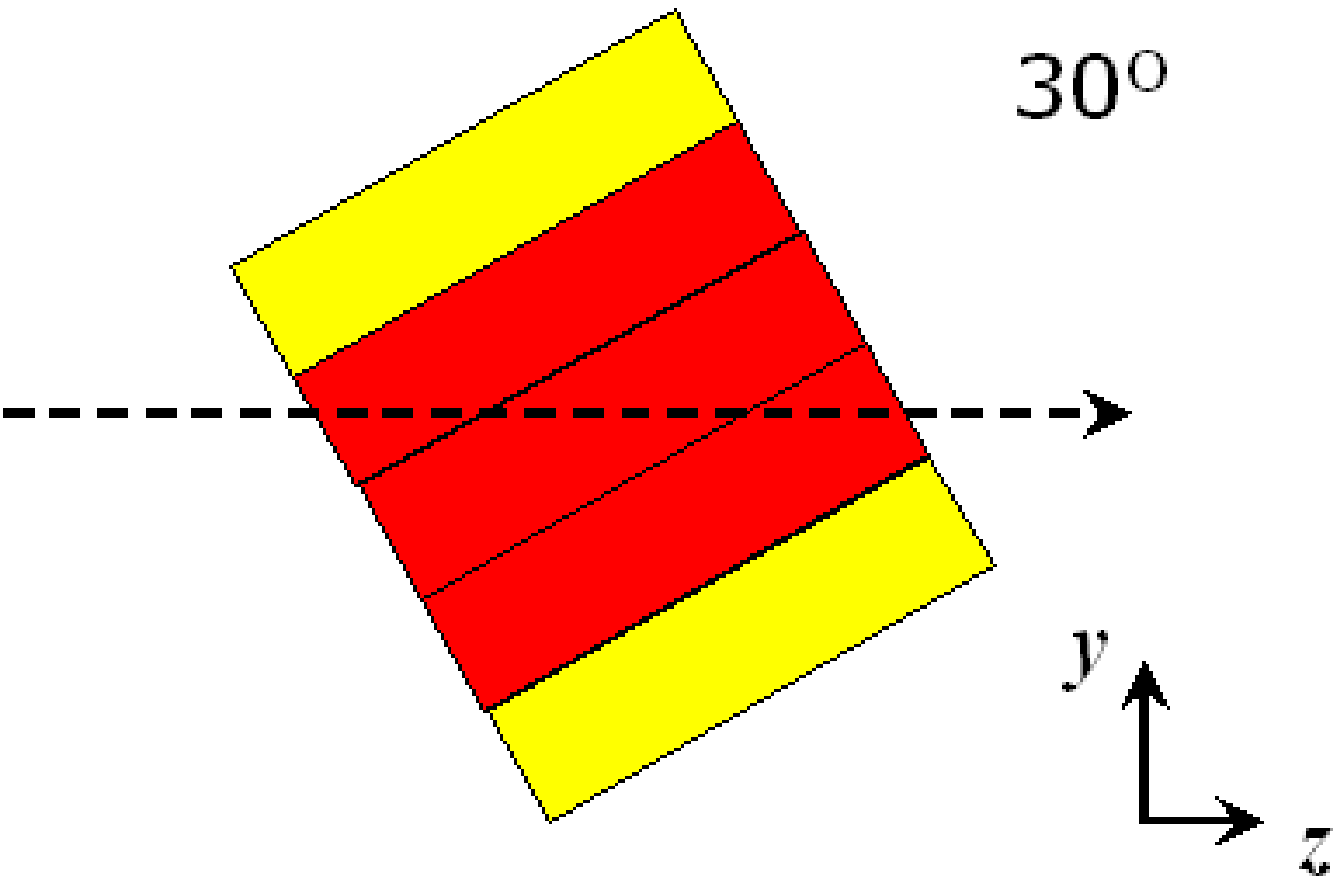}
\end{center}  
\caption{}
\label{figtiltangles02}
\end{figure}

\begin{figure}[hb!]
\centering
\includegraphics[scale=0.35]{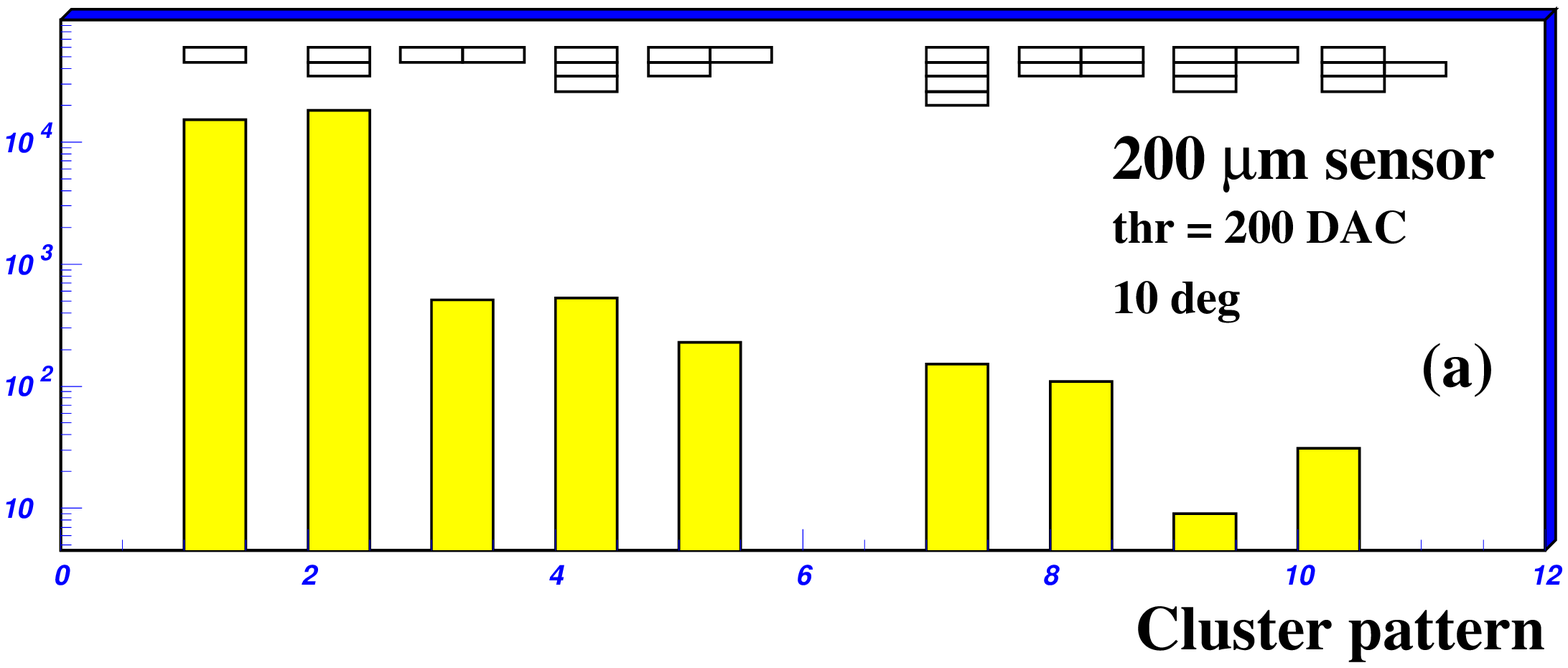}
\\
\includegraphics[scale=0.35]{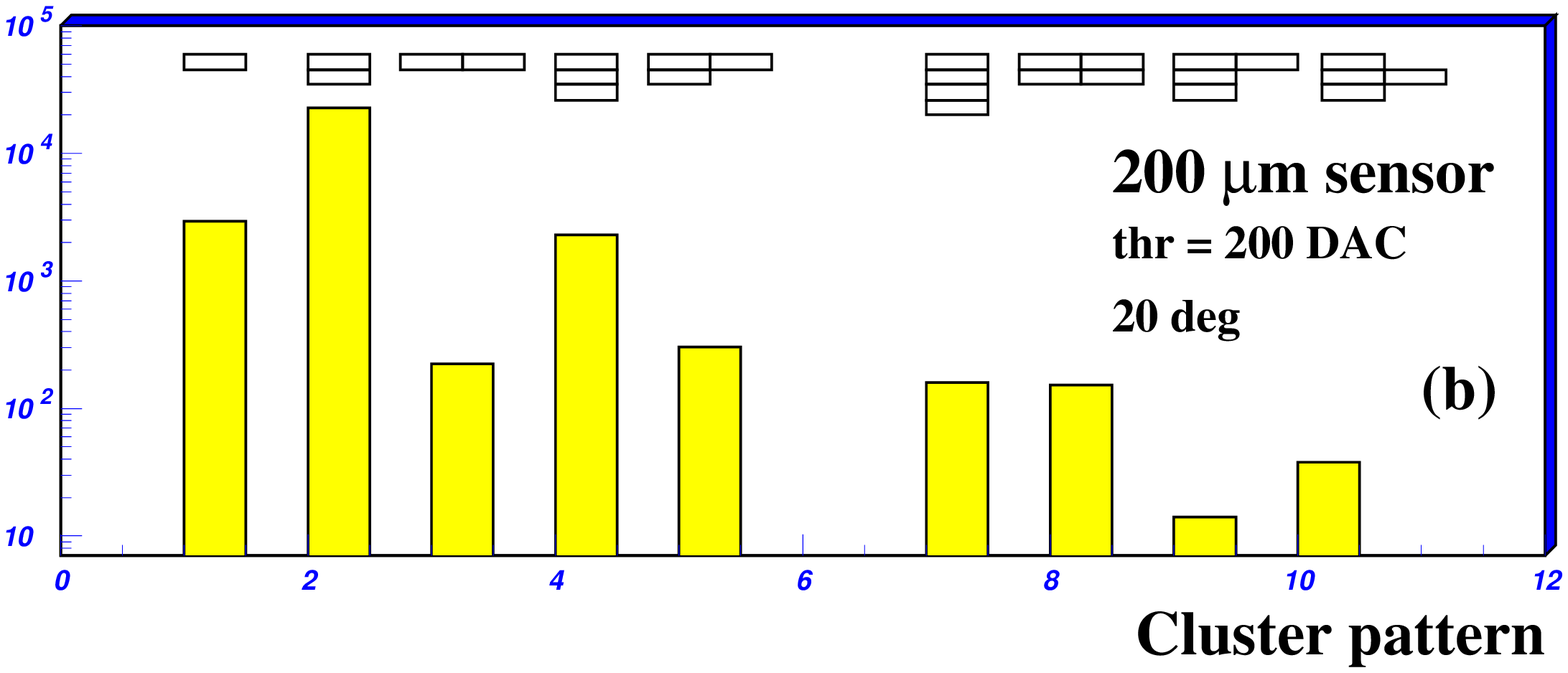}
\caption{}
\label{figclustertype02_tilt10-20_th200_log}
\end{figure}

\begin{figure}[ht!]
\includegraphics[scale=0.5]{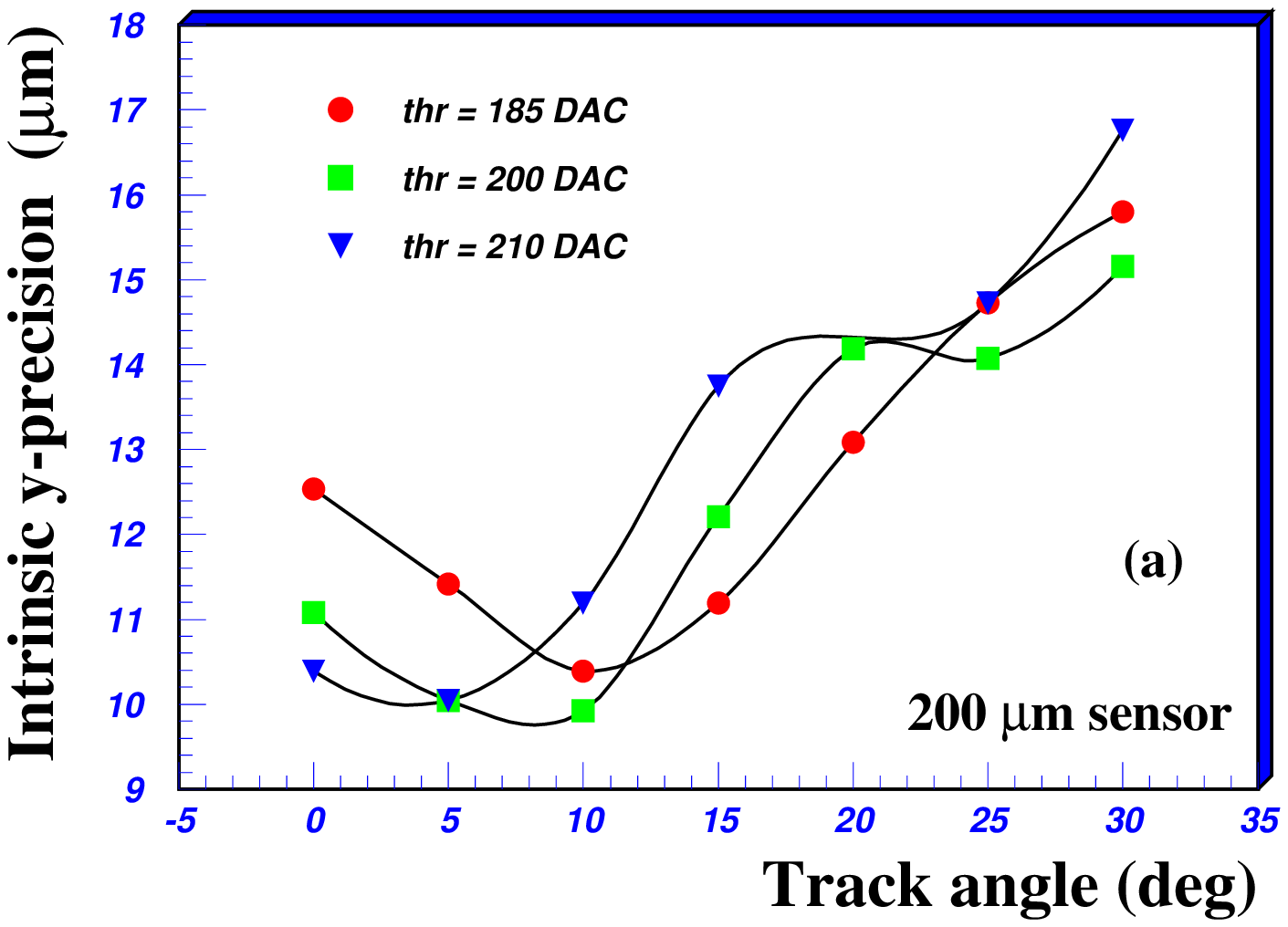}
\hspace{\fill}
\begin{minipage}[t]{110mm}
\includegraphics[scale=0.5]{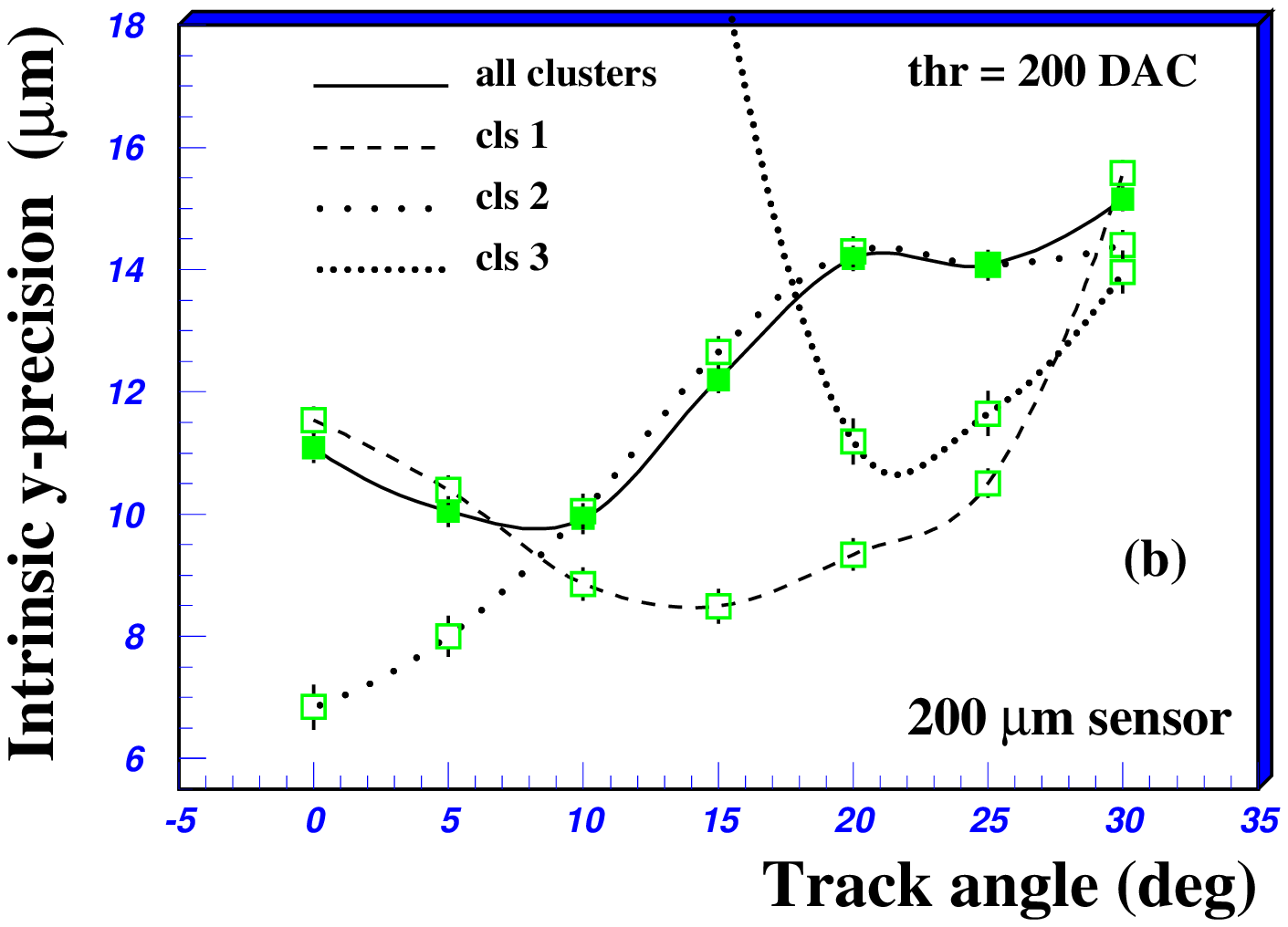}
\end{minipage}
\vspace{-9mm}
\caption{}
\label{figresolutionsvsang02}
\end{figure}

Data runs were taken with the detector under
test tilted from 0 to 30 
degrees and with thresholds set to 185, 200 and 210 DAC units.
Fig.\ref{figtiltangles02} schematically shows
pixel cells traversed by a track at three different incidence angles.
As an example of the evolution of the cluster patterns by varying
the track inclination angle,
in Fig.\ref{figclustertype02_tilt10-20_th200_log} we report
the corresponding distributions for 10 (a)
and 20 degrees (b), to be compared with that
shown in Fig.\ref{figclustertype02_tilt0_th200_log} for
normal track incidence. For 10 degrees inclined tracks the two
main cluster patterns are almost balanced, while 
at 20 degrees the double pixel clusters dominate.
\\
Results of the study on the spatial precision
as a function of the track incidence angles
are shown in Fig.\ref{figresolutionsvsang02}.
The precision curves reach the
minimum around 5-10 degrees
(equal fractions of single
and double-$y$ pixel clusters) then degrading for all the threshold 
settings with the increasing track angle.
In Fig.\ref{figresolutionsvsang02}(b) the intrinsic
precision is shown together with the
contributions due to the main pixel cluster patterns separately, for the case
of 200 DAC threshold setting.
These results can be particularly useful in the tuning of the
tracking errors to be associated to the SPD points, both in
the simulation and in the data analysis.

\section{Summary}
The performance of prototype assemblies for the 
ALICE Silicon Pixel Detector have been extensively studied by using 
beam test data collected in the past years at the CERN SPS.
The cluster pattern distribution, the intrinsic spatial
precision and the detection efficiency 
have been investigated as a function
of both the detector threshold and the incident angle of
the tracks.
The results show a very high detection efficiency (above 99\%)
in a wide threshold range and
a spatial precision
of about 10 $\mu{m}$ in the short pixel side coordinate
for normal track incidence and 210 DAC threshold. The
detector performance with angled tracks has also been
investigated.

\newpage
%\vspace{2.5cm}

\noindent{\bf Figure caption}
\vspace{5mm}

%ccc
\setcounter{figcount}{0}
\stepcounter{figcount}

Fig. \thefigcount. Schematic view of the layout for the 2002 ALICE SPD beam test.
\stepcounter{figcount}
\vspace{7mm}
\\
Fig. \thefigcount. Distribution of cluster patterns on the test plane, for
tracks at normal incidence and threshold setting of 200 DAC units.
\stepcounter{figcount}
\vspace{7mm}
\\
Fig. \thefigcount. Distributions of the $y$-residuals for 
all the pixel clusters (a) and for double-$y$ pixel clusters only (b);
$\sigma_{resid}$ is the standard deviation of the Gaussian fit to
each distribution.
\stepcounter{figcount}
\vspace{7mm}
\\
Fig. \thefigcount. Detection efficiency as a function of the threshold for
tracks at normal incidence, for both sensor thicknesses. The
results of a fit with Gaussian-integral function are superimposed.
\stepcounter{figcount}
\vspace{7mm}
\\
Fig. \thefigcount. Intrinsic spatial precision along the $y$ coordinate for tracks
at normal incidence, as a function of the threshold and for 
both sensor thicknesses.
\stepcounter{figcount}
\vspace{7mm}
\\
Fig. \thefigcount. Contribution of single and double-$y$ pixel clusters
to the intrinsic precision (a) and to the detector efficiency (b)
as a function of the threshold. 
\stepcounter{figcount}
\vspace{7mm}
\\
Fig. \thefigcount. Schematic picture of pixel planes crossed by tracks at
different inclination angles in the $yz$ planes.
Pixel cells traversed by the track are shown in red.
\stepcounter{figcount}
\vspace{7mm}
\\
Fig. \thefigcount. Cluster pattern distributions for threshold settings of
200 DAC units, for tracks at 10 (a) and 20 degrees (b) incidence angle.
\stepcounter{figcount}
\vspace{7mm}
\\
Fig. \thefigcount. Intrinsic precision in the $y$ coordinate as a
function of the track incidence angle on the detector and
at different threshold settings. 
\stepcounter{figcount}
\vspace{7mm}
\\
Fig. \thefigcount. 
Intrinsic $y$-precision as a
function of the track incidence angle,
at different thresholds (a) and at 200 DAC threshold
for the main cluster patterns separately (b).

\newpage

%fff

\vspace{10mm} 

\vspace{10mm} 

\newpage

\vspace{5mm}

\vspace{5mm}

\newpage

\vspace{5mm}

\vspace{5mm}

\end{document}